\documentclass[letterpaper, 10 pt, conference]{ieeeconf}  

\IEEEoverridecommandlockouts                              
\usepackage[utf8]{inputenc}
\usepackage[T1]{fontenc}
\usepackage{graphicx}
\usepackage{cite}
\usepackage{epstopdf}
\usepackage{hyperref}
\usepackage{amsfonts}
\usepackage{subcaption}
\usepackage{amssymb}
\usepackage{amsmath}
\usepackage[capitalise]{cleveref}
\usepackage{mathtools}
\usepackage{breqn}
\usepackage{wrapfig}
\usepackage{xcolor}
\usepackage[percent]{overpic} 
\usepackage{float}
\usepackage{cite}
\usepackage{bm}
\DeclareMathAlphabet      {\mathbfit}{OML}{cmm}{b}{it}

\graphicspath{{Figures/}}

\title{\LARGE \bf
Modeling of Non-linear Dynamics of Lithium-ion Batteries via Delay-Embedded Dynamic Mode Decomposition
}

\author{Khalid Mahmud Labib and Shabbir Ahmed$^{1}$\\
Dynamical Signals and Systems Lab (DSSL)\\ Department of Mechanical Engineering\\
South Dakota State University, Brookings, SD 
\thanks{$^{1}$Corresponding author; Email {\tt\small shabbir.ahmed@sdstate.edu}}.%
}

\begin{document}

\maketitle
\thispagestyle{empty}
\pagestyle{empty}

\begin{abstract}

The complex electrochemical behavior of lithium-ion batteries results in non-linear dynamics and appropriate modeling of this non-linear dynamical system is of interest for better management and control. In this work, we proposed a family of dynamic mode decomposition (DMD)-based data-driven models that do not require detailed knowledge of the composition of the battery materials but can essentially capture the non-linear dynamics with higher computational efficiency. Only voltage and current data obtained from hybrid pulse power characterization (HPPC) tests were utilized to form the state space matrices and subsequently used for predicting the future terminal voltage at different state of charge (SoC) and aging levels. To construct the system model, 60\% of the data from a single HPPC test was utilized to generate time-delay embedded snapshots, with embedding dimension ranging from 40 to 2000. Among these, an embedding dimension of 1810 resulted in the least residual sum of squares (RSS) error of 3.86 for the dynamic mode decomposition with control (DMDc) model and 30 for the standard DMD model. For DMDc model, delay embeddings (ranging from 1 to 12) were also incorporated into the input current signals. For the input matrix, an embedding dimension of 6 resulted in a minimum RSS error of 1.74. Furthermore, the system matrices A and B, identified from the HPPC test when the cell is in its healthy state, were held fixed and used to simulate the system dynamics for aged batteries by updating only the control input. Despite the presence of nonlinear degradation effects in later cycles, the DMDc model effectively captured key inner dynamics such as voltage dips and transient responses for subsequent charge and discharge cycles. 



\end{abstract}

\section{Introduction} \label{sec:intro}

Lithium-ion batteries are widely used in modern energy storage systems due to their high energy density, high power capability, and long service life \cite{ahmed2025autoregressive}. They play a vital role in applications ranging from electric vehicles to aircraft power systems \cite{FARHANGDOUST2023}. Thus, effective health management of these batteries is essential for the reliable and safe operation of the associated systems. To achieve this, battery management systems (BMS) are integrated into battery packs, which monitor and predict critical battery indicators such as state of charge (SoC) and state of health (SoH) \cite{Xiong2018,Wang2020}. This requires the formulation of battery models that are faithful to the actual dynamics of the battery system and can track the progression of degradation over time and usage. 

Battery modeling approaches can be broadly classified into three main categories: physics-based models (developed from electrochemical and physical first principles), empirical equivalent circuit models (ECM), and data-driven models. Among all the different types of models, ECMs are the most widely used models to analyze battery dynamics \cite{lai2018comparative,quelin2023coupling} and have been integrated into the battery management systems. In an equivalent circuit model, the complex electrochemical processes within a battery are represented by a simplified electrical circuit composed of resistors, capacitors, and voltage elements. These components are arranged to predict the behavior of a physical lithium-ion battery. Although a battery does not contain these electronic components internally, its voltage response (output) to an input current (excitation) is similar to that of an equivalent circuit model for the same current input. As a result, these models are extensively used in real-time control algorithms in commercial battery management systems. \cite{tran2021comprehensive,petri2023state}. However, the precision of equivalent circuit models (ECM) is significantly influenced by the choice of circuit topology and the availability of precise battery characterization data. Since batteries exhibit complex nonlinear behavior, developing a reliable equivalent circuit model requires careful parameterization and consideration of the aging phenomena or battery degradation. Due to these constraints, data-driven approaches have gained favor because of their robustness to capture complicated nonlinear dynamics without requiring deep physical knowledge or extensive battery characterization. These approaches may employ machine learning algorithms to learn the input-output relationships, such as mapping the state of charge, temperature, and current to voltage and capacity directly from large datasets. Although these models have been successful in many cases, however, they may lack the explainability and interpretability of equivalent circuit models.

Previously, sparse identification of non-linear dynamics (SINDy) and non-linear autoregressive with exogenous input (NARX) models have been employed to extract governing equations directly from data \cite{kaiser2018sparse,rudy2017data,piroddi2003identification,ahmed2021stochastic}. However, SINDy and NARX-type models may not readily return the state transition and input matrices, typically known as $A$ and $B$ matrices, which form the basis of state-space control and filtering, such as linear quadratic regulator, model predictive control, and Kalman filtering. Additionally, SINDy approach may need time derivative and may become noise sensitive. In these contexts, dynamic mode decomposition (DMD) and dynamic mode decomposition with control (DMDc) algorithms may perform better as they employ singular value decomposition (SVD) as the underlying data processing algorithm, which is well studied and mathematically robust \cite{proctor2016dynamic,muld2012flow,pan2011dynamical}. As a result, DMD can handle large state vector and data such as images, measurements from dense sensor arrays etc. efficiently, whereas SINDy and NARX may face challenges \cite{belhumeur1997eigenfaces,swets2002using}. Additionally, DMD models are linear and amenable to state-space representation, directly providing the state transition and input matrices. Explainability and representation parsimony may be obtained from the modes and poles extracted from the state-transition matrix $A$. Prior works have included state of charge (SoC) \cite{AbuSeif2022} and capacity as states in the formulation of the input and output matrices, however, SoC and capacity are derived quantities, which may not be readily available and need to be estimated \cite{petri2023state,severson2019data,xing2014state}. Hence, in this study, we have used only the voltage information as the available measurements or outputs and explored the prediction of the non-linear dynamics with the linear DMD formulation.

The paper is structured as follows: Section 2 outlines the mathematical formulations of the Hankel matrix, DMD, and DMDc. Section 3 describes the experiment and data generation, and Section 4 presents the findings from this study and discusses its implications. Finally, Section 5 summarizes the major findings and provides direction about the future works.

\section{Background}

This section outlines the mathematical formulation of the Hankel matrix, Dynamic Mode Decomposition (DMD), and Dynamic Mode Decomposition with control (DMDc) algorithms, and how these algorithms have been applied in the context of modeling of lithium-ion battery dynamics.

\subsection{Hankel Matrix}

A Hankel matrix is a special type of matrix where all elements along each anti-diagonal are constant. It is widely used in system identification, signal processing, and control theory. When constructed from a sequential data set, they transform a one-dimensional signal into a higher-dimensional representation. From a sequence of discrete signal $[x(1), x(2), \dots, x(N)]$, a Hankel matrix can be constructed as follows:
\[
\mathbf{H} =
\begin{bmatrix}
x(1) & x(2) & \cdots & x({n}) \\
x(2) & x(3) & \cdots & x({n+1}) \\
\vdots & \vdots & \ddots & \vdots \\
x({m}) & x({m_1}) & \cdots & x(N)
\end{bmatrix}
\]
where $1 < m < N$, $n = N - m + 1$, and $H \in \mathbb{R}^{m \times n}$, with $m$ denoting the embedding dimension and $n$ denoting number of columns. By choosing an appropriate embedding dimension $m$ (and delay $\tau$), a block Hankel (time-delay) matrix built from $m$ consecutive delayed samples encodes temporal dependencies in the data. This embedding can help the approximation of the underlying dynamics-linear or non-linear, when used with appropriate identification algorithms.


\subsection{Dynamic Mode Decomposition}

Dynamic Mode Decomposition (DMD) is a data-driven method that identifies a linear operator, which best advances the snapshot data forward in time, yielding spatiotemporal modes and their growth or decay rates. In this study, we formed snapshots from the voltage time series data by constructing a Hankel matrix-based time-delay embedding, where each column stacks $m$ delayed samples from the hybrid pulse power characterization (HPPC) dataset ($[v(t_1), v(t_2), \dots, v(t_N), v(t_{N+1})]$) and then apply DMD to the resulting snapshot matrices. The resulting snapshot matrices are denoted by $\mathbf{X}$ and $\mathbf{X}'$: \\


\begin{equation}\label{eq:x_mat}
\mathbf{X} =
\begin{bmatrix}
v(t_{1}) & v(t_{2}) & \cdots & v(t_{n}) \\
v(t_{2}) & v(t_{3}) & \cdots & v(t_{n+1}) \\
\vdots    & \vdots   & \ddots & \vdots \\
v(t_{m})  & v(t_{m+1}) & \cdots & v(t_{N})
\end{bmatrix}
\end{equation}

\begin{equation}\label{eq:x_mat2}
\mathbf{X}' =
\begin{bmatrix}
v(t_{2}) & v(t_{3}) & \cdots & v(t_{n+1}) \\
v(t_{3}) & v(t_{4}) & \cdots & v(t_{n+2}) \\
\vdots    & \vdots   & \ddots & \vdots \\
v(t_{m+1}) & v(t_{m+2}) & \cdots & v(t_{N+1})
\end{bmatrix}
\end{equation}

The DMD algorithm estimates a linear operator $\mathbf{A}$, in the least squares sense, that advances the snapshot matrix one step ahead
\begin{equation}
    \mathbf{X'}=\mathbf{A}\mathbf{X}
\end{equation}

The best-fit operator $\mathbf{A}$ thus defines a transition matrix of a linear dynamical system that optimally advances the snapshots forward in time. With uniform sampling, the high-dimensional state vector $\mathbf{x} \in \mathbb{R}^{n}$ can be assumed to progress in the following way:

\begin{equation}
    \mathbf{x}_{k+1} = \mathbf{A}\mathbf{x}_{k}
    \label{eq:model}
\end{equation}

where, $\mathbf{A}$ is given by:
\begin{equation}
    \mathbf{A}=\mathbf{X'}\mathbf{X^{\dagger}}
    \label{eq:amatrix}
\end{equation}

Here, $\mathbf{X}^\dagger$ denotes the Moore–Penrose pseudoinverse of the matrix $\mathbf{X}$, and $\mathbf{A} \in \mathbb{R}^{n \times n}$. The best-fit operator $\mathbf{A}$ minimizes the Frobenius norm $|\mathbf{X}' - \mathbf{A}\mathbf{X}|_F$. A computationally efficient and accurate approach to compute the pseudoinverse is through the singular value decomposition (SVD). The SVD of $\mathbf{X}$ yields the well-known factorization:
\begin{equation}
    \mathbf{X}=\mathbf{U}{\mathbf{\Sigma}}{\mathbf{V}}^{*}
\end{equation}
where $\mathbf{U} \in \mathbb{C}^{n \times r}$, $\mathbf{\Sigma} \in \mathbb{C}^{r \times r}$, and $\mathbf{V} \in \mathbb{C}^{m \times r}$. Typically, $r \leq m$. An appropriate value of $r$ needs to be used during computation. $\mathbf{V}^{*}$ represents the conjugate transpose of $\mathbf{V}$.  Using the SVD of the snapshot matrix $\mathbf{X}$, the $\mathbf{A}$ can be approximated from Equation \ref{eq:amatrix} as:

\begin{equation}
    \mathbf{A} = \mathbf{X'}\mathbf{V} \, \mathbf{\Sigma}^{-1} \mathbf{U}^{*}
    \label{eq:A_comp}
\end{equation}

Once $\mathbf{A}$ is computed from Equation \ref{eq:A_comp}, a dynamic model of the system can be constructed from Equation \ref{eq:model}. Note that Equation \ref{eq:model} does not use any input or perturbation.

\subsection{Dynamic Mode Decomposition with Control}

This section formulates dynamic mode decomposition with control (DMDc), which estimates a low-rank linear, time-invariant model directly from measurements and known inputs.

The DMDc method aims to identify the best-fit linear operators $\mathbf{A}$ and $\mathbf{B}$ that approximately satisfy the following dynamical relationship based on the measurement data:

\begin{equation}
    \mathbf{x}_{k+1} = \mathbf{A}\mathbf{x}_{k} + \mathbf{B}\mathbf{u}_{k}
    \label{eq:model_dmdc}
\end{equation}


In addition to the snapshot matrices $\mathbf{X}$ and $\mathbf{X}'$ (Equation \ref{eq:x_mat} and \ref{eq:x_mat2}, respectively ), a Hankel matrix representing the actuation or input history can also be constructed as follows\\

$$
\mathbf{U} = \left[ \begin{array}{cccc}
u(t_1) & u(t_2) & \cdots & u(t_{n}) \\
u(t_2) & u(t_3) & \cdots & u(t_{n+1}) \\
\vdots & \vdots & \ddots & \vdots \\
u(t_{m}) & u(t_{m+1}) & \cdots & u(t_N)
\end{array} \right]
$$

The dynamics in \eqref{eq:model_dmdc} can be expressed in terms of data matrices as
\begin{equation}\label{eq:dmdc_data}
    \mathbf{X}' \approx \mathbf{A}\,\mathbf{X} + \mathbf{B}\,\mathbf{U},
\end{equation}
where \(\mathbf{X}=[\,\mathbf{x}_1,\ldots,\mathbf{x}_{N}\,]\) and \(\mathbf{X}'=[\,\mathbf{x}_2,\ldots,\mathbf{x}_{N+1}\,]\) collect measurement snapshots, and
\(\mathbf{U}=[\,\mathbf{u}_1,\ldots,\mathbf{u}_{n}\,]\) collects the corresponding input snapshots.
This formulation is advantageous for experimental studies because it requires only measured outputs and known inputs to estimate the intrinsic operator \(\mathbf{A}\) and the input map \(\mathbf{B}\).

Stacking \(\mathbf{X}\) and \(\mathbf{U}\) gives
\begin{equation}\label{eq:stacked}
    \boldsymbol{\Omega} \;:=\;
    \begin{bmatrix}
        \mathbf{X}\\[2pt]
        \mathbf{U}
    \end{bmatrix},
    \qquad
    \mathbf{G} \;:=\; \begin{bmatrix}\mathbf{A} & \mathbf{B}\end{bmatrix},
    \qquad
    \mathbf{X}' \approx \mathbf{G}\,\boldsymbol{\Omega}.
\end{equation}
The least-squares estimate is therefore
\begin{equation}\label{eq:ls_solution}
    \mathbf{G} \;\approx\; \mathbf{X}'\,\boldsymbol{\Omega}^{\dagger},
\end{equation}
where \(^{\dagger}\) denotes the Moore--Penrose pseudoinverse.

Let the rank-\(r\) SVD of \(\boldsymbol{\Omega}\) be
\begin{equation}\label{eq:svd_omega}
    \boldsymbol{\Omega} \;\approx\; \tilde{\mathbf{U}}_r\,\tilde{\boldsymbol{\Sigma}}_r\,\tilde{\mathbf{V}}_r^{*},
    \qquad
    \boldsymbol{\Omega}^{\dagger} \;\approx\; \tilde{\mathbf{V}}_r\,\tilde{\boldsymbol{\Sigma}}_r^{-1}\,\tilde{\mathbf{U}}_r^{*}.
\end{equation}
Here \(\tilde{\boldsymbol{\Sigma}}_{r}\) holds the \(r\) dominant singular values. By performing truncation, one can denoise the data and avoid inverting tiny singular values. By partitioning the left singular vectors conformably with the stacked rows, the below representation can be obtained
\begin{equation}\label{eq:U_partition}
    \tilde{\mathbf{U}}_r \;=\;
    \begin{bmatrix}
        \tilde{\mathbf{U}}_{x}\\[2pt]
        \tilde{\mathbf{U}}_{u}
    \end{bmatrix},
\end{equation}
where \(\tilde{\mathbf{U}}_{x}\) spans the output (measurement) subspace and \(\tilde{\mathbf{U}}_{u}\) spans the input subspace.

Similarly, a rank-\(r_x\) SVD of \(\mathbf{X}'\) can be taken to define an output projection basis:
\begin{equation}\label{eq:svd_xprime}
    \mathbf{X}' \;\approx\; \hat{\mathbf{U}}_{r_x}\,\hat{\boldsymbol{\Sigma}}_{r_x}\,\hat{\mathbf{V}}_{r_x}^{*}.
\end{equation}
Projecting \eqref{eq:ls_solution} onto \(\hat{\mathbf{U}}_{r_x}\) yields a reduced operator as follows:
\begin{equation}\label{eq:G_tilde}
    \tilde{\mathbf{G}}
    \;:=\;
    \hat{\mathbf{U}}_{r_x}^{*}\,\mathbf{G}
    \;\approx\;
    \hat{\mathbf{U}}_{r_x}^{*}\,\mathbf{X}'\,\tilde{\mathbf{V}}_r\,\tilde{\boldsymbol{\Sigma}}_r^{-1}\,\tilde{\mathbf{U}}_r^{*}
    \;=\;
    \begin{bmatrix}\tilde{\mathbf{A}} & \tilde{\mathbf{B}}\end{bmatrix}.
\end{equation}
Extracting the reduced matrices uses the partition \eqref{eq:U_partition}:
\begin{subequations}\label{eq:AB_tilde}
\begin{align}
    \tilde{\mathbf{A}}
    &= \hat{\mathbf{U}}_{r_x}^{*}\,\mathbf{A}\,\hat{\mathbf{U}}_{r_x}
     \;\approx\;
     \hat{\mathbf{U}}_{r_x}^{*}\,\mathbf{X}'\,\tilde{\mathbf{V}}_r\,\tilde{\boldsymbol{\Sigma}}_r^{-1}\,
     \tilde{\mathbf{U}}_{x}^{*}\,\hat{\mathbf{U}}_{r_x}, \label{eq:A_tilde}\\[4pt]
    \tilde{\mathbf{B}}
    &= \hat{\mathbf{U}}_{r_x}^{*}\,\mathbf{B}
     \;\approx\;
     \hat{\mathbf{U}}_{r_x}^{*}\,\mathbf{X}'\,\tilde{\mathbf{V}}_r\,\tilde{\boldsymbol{\Sigma}}_r^{-1}\,
     \tilde{\mathbf{U}}_{u}^{*}. \label{eq:B_tilde}
\end{align}
\end{subequations}

Once \(\mathbf{A}\) and \(\mathbf{B}\) (or their reduced counterparts) are obtained, the one-step ahead prediction of the dynamical system can be approximated by
\begin{equation}\label{eq:state_update}
    \mathbf{x}_{k+1} \;\approx\; \mathbf{A}\,\mathbf{x}_k \;+\; \mathbf{B}\,\mathbf{u}_k .
\end{equation}


\section{Experimental set-up}


A 30 Ah lithium-ion battery was employed to generate hybrid pulse power characterization (HPPC) test data for the purpose of modeling and analyzing the battery's dynamic behavior. The battery operated within a voltage range of 2.5 V to 4.2 V. Initially, the battery was fully charged using a two-step charging protocol consisting of: (1) constant current (CC) charging, followed by (2) constant voltage (CV) charging. Subsequently, the battery was allowed to rest for one hour (Step 3). A 10 A constant current discharge pulse was then applied for 10 seconds (Step 4), after which the battery rested for 3 minutes (Step 5). This was followed by a 5 A constant current charge pulse applied for 20 seconds (Step 6), and another rest period of 2 minutes (Step 7). Next, a 10 A constant current discharge was carried out for 18 minutes, corresponding to a 0.1 Ah capacity reduction, terminating when the battery voltage reached 3.95 V. After this, the battery was rested for 1 hour, during which the voltage gradually recovered to approximately 4.07 V (Step 8). Steps 3 through 8 were repeated 10 times to simulate various state of charge (SoC). Following the completion of the HPPC test on a fresh (healthy) battery, the cell was subjected to continuous cycling to induce capacity degradation. After every 20 charge-discharge cycles, an HPPC test (using the same procedure as described above) was conducted to assess changes in the battery's voltage response, thereby characterizing the effects of degradation over time.

\begin{figure}[t]
    \centering
    \begin{subfigure}[b]{\columnwidth}
        \centering
        \includegraphics[width=\linewidth]{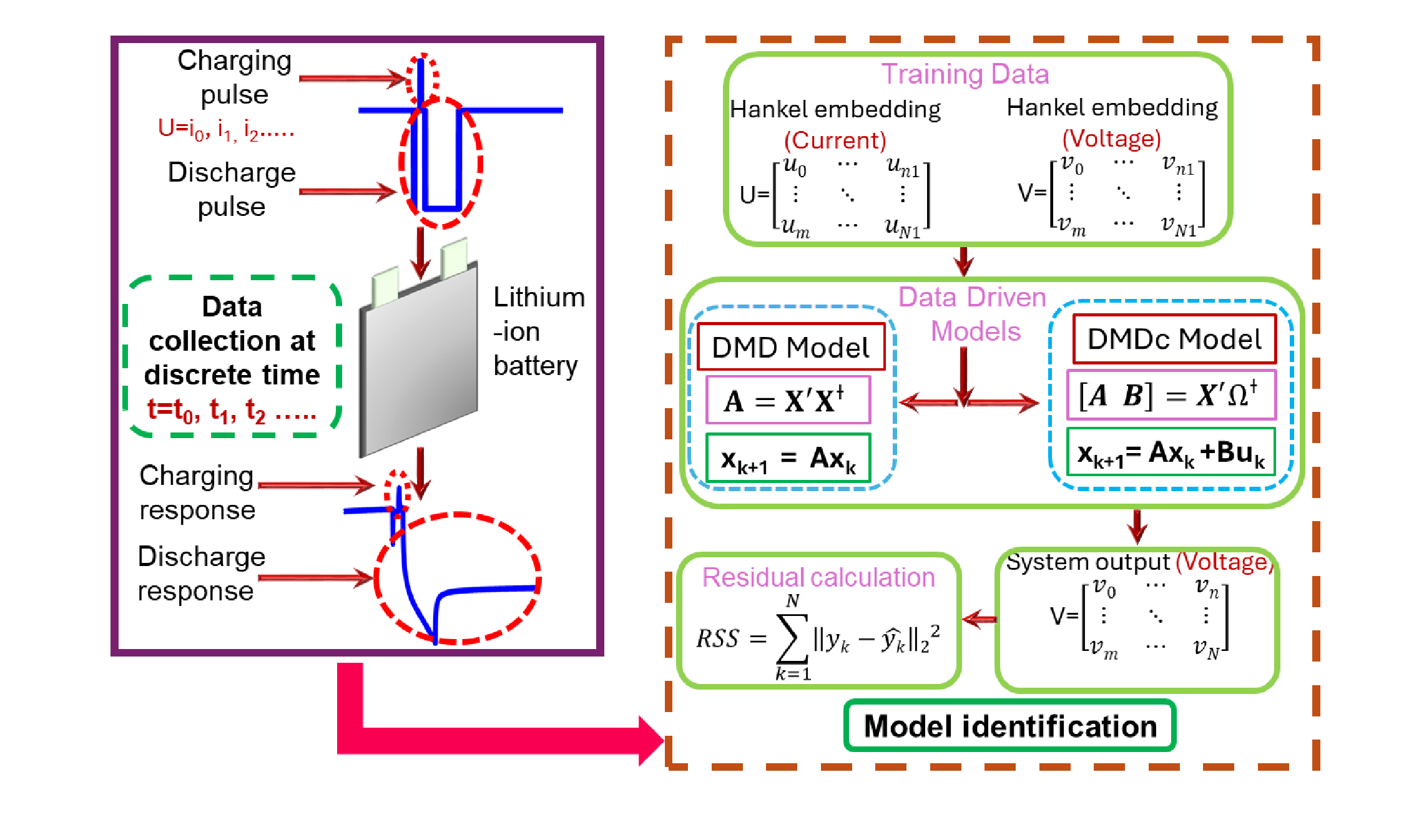}
        
    \end{subfigure}
    
    \caption{Schematic representation of the experimental setup and modeling framework}
    \label{fig:all cycle}
\end{figure}

%
\section{Results and Discussions}


In this section, we report simulation results for dynamic mode decomposition (DMD) and dynamic mode decomposition with control (DMDc). We describe model identification (rank/embedding selection and least-squares fitting) and present predictions of charge-discharge behavior, followed by an analysis of the identified dynamics. 

For experimental validation, we compare the model-predicted or simulated charge-discharge profiles with measured profiles under the same input sequences. Predictive accuracy is quantified using the residual sum of squares (RSS),
\begin{equation}
    \mathrm{RSS} \;=\; \sum_{k=1}^{N} \bigl\| \mathbf{y}_k - \hat{\mathbf{y}}_k \bigr\|_2^2,
\end{equation}
where \(\mathbf{y}_k\) and \(\hat{\mathbf{y}}_k\) denote the measured and predicted outputs at time step \(k\), respectively (for scalar outputs, \(\|\cdot\|_2^2\) reduces to a squared difference).

\subsection{Formation of Hankel Matrix}

In standard DMD, the snapshot matrix is typically assembled directly from measured outputs. In our setting only a single output (voltage) is available, and the true state is unobserved, which can limit the model’s ability—especially for nonlinear battery dynamics—to capture the underlying behavior. To enrich the observable coordinates, we construct a block Hankel (time–delay) embedding of the voltage time series, lifting temporal structure into an \(m\)-dimensional space. Concretely, letting \(y_k\) denote voltage, we form \(\mathbf{X}=\mathcal{H}_m(y)\), whose columns stack \(m\) consecutive delayed samples.

\begin{figure}[t!]
    \centering
    \begin{picture}(200,290)
    \put(-20,150){ \includegraphics[width=0.93\columnwidth]{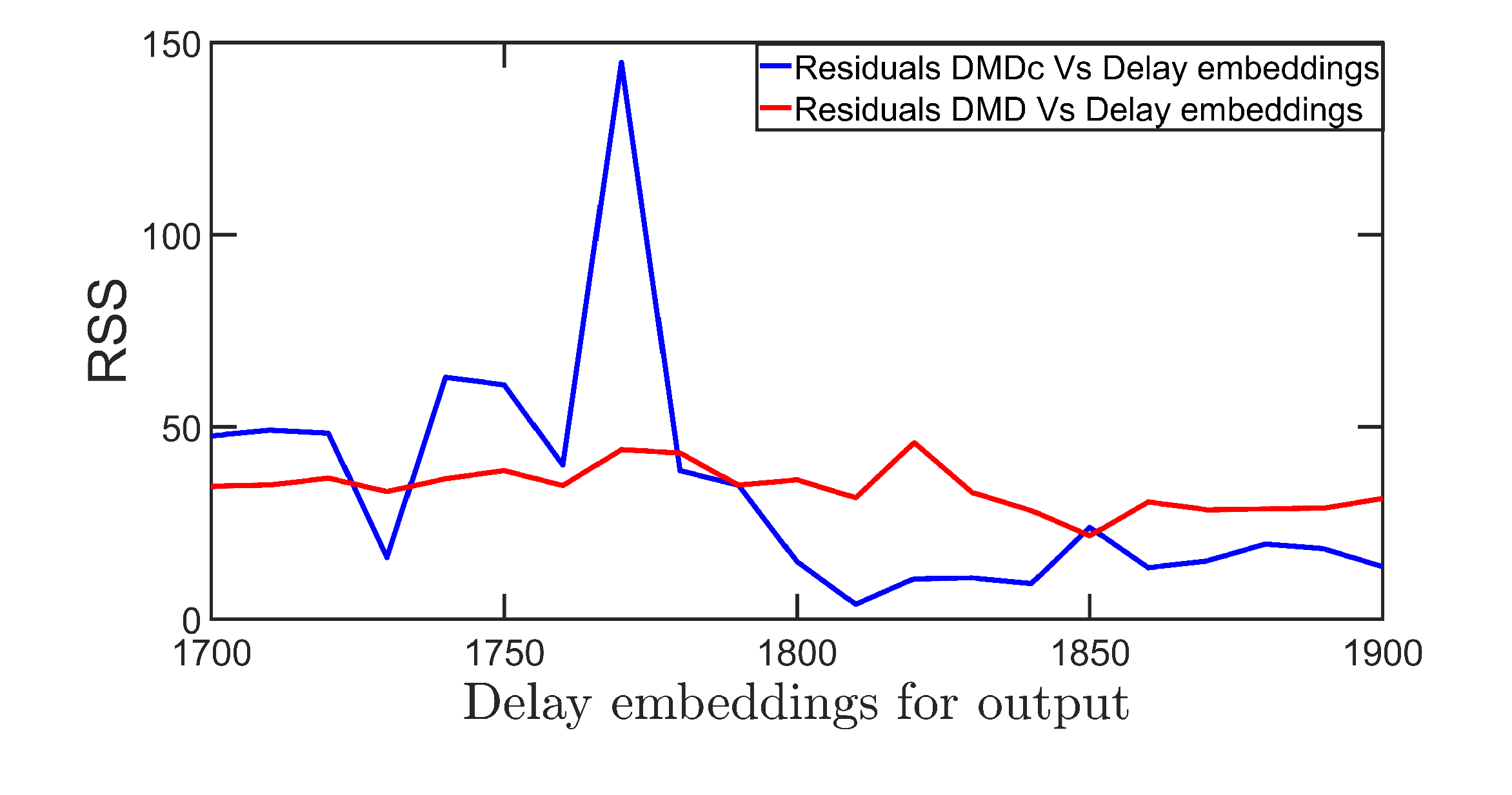}}
    \put(-10,5){\includegraphics[width=0.9\columnwidth]{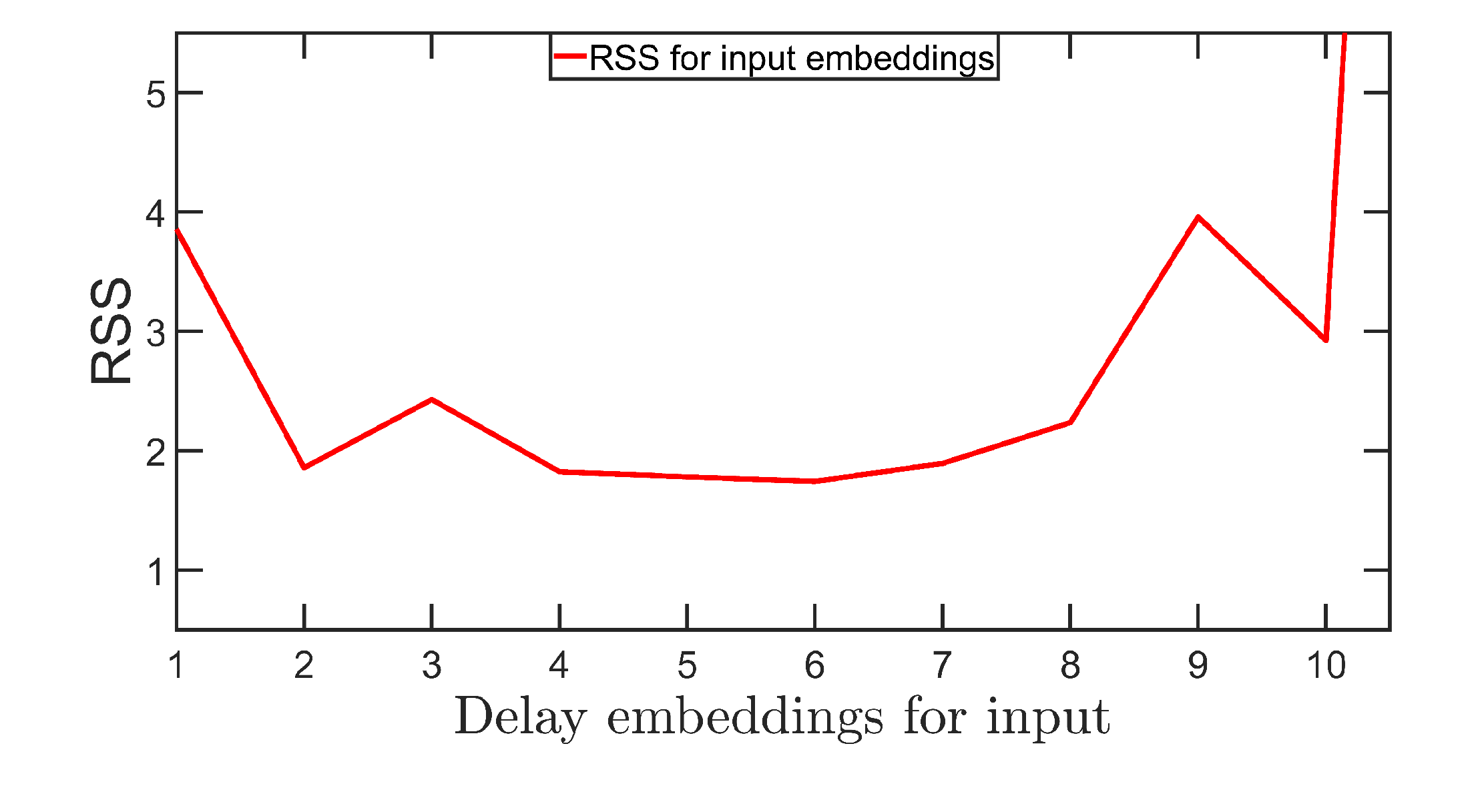}}

    \put(10,250){\large \textbf{(a)}}
    \put(10,105){ \large \textbf{(b)}}
    \end{picture}
    
    \caption{Comparison of RSS for different delay embeddings (a) Voltage(output) (b) current(input) in DMD and DMDc models.} 
\label{fig:RSS1} 
\end{figure}

\begin{figure}[H]
    \centering
    \begin{subfigure}[b]{\columnwidth}
        \centering
        \includegraphics[width=\linewidth]{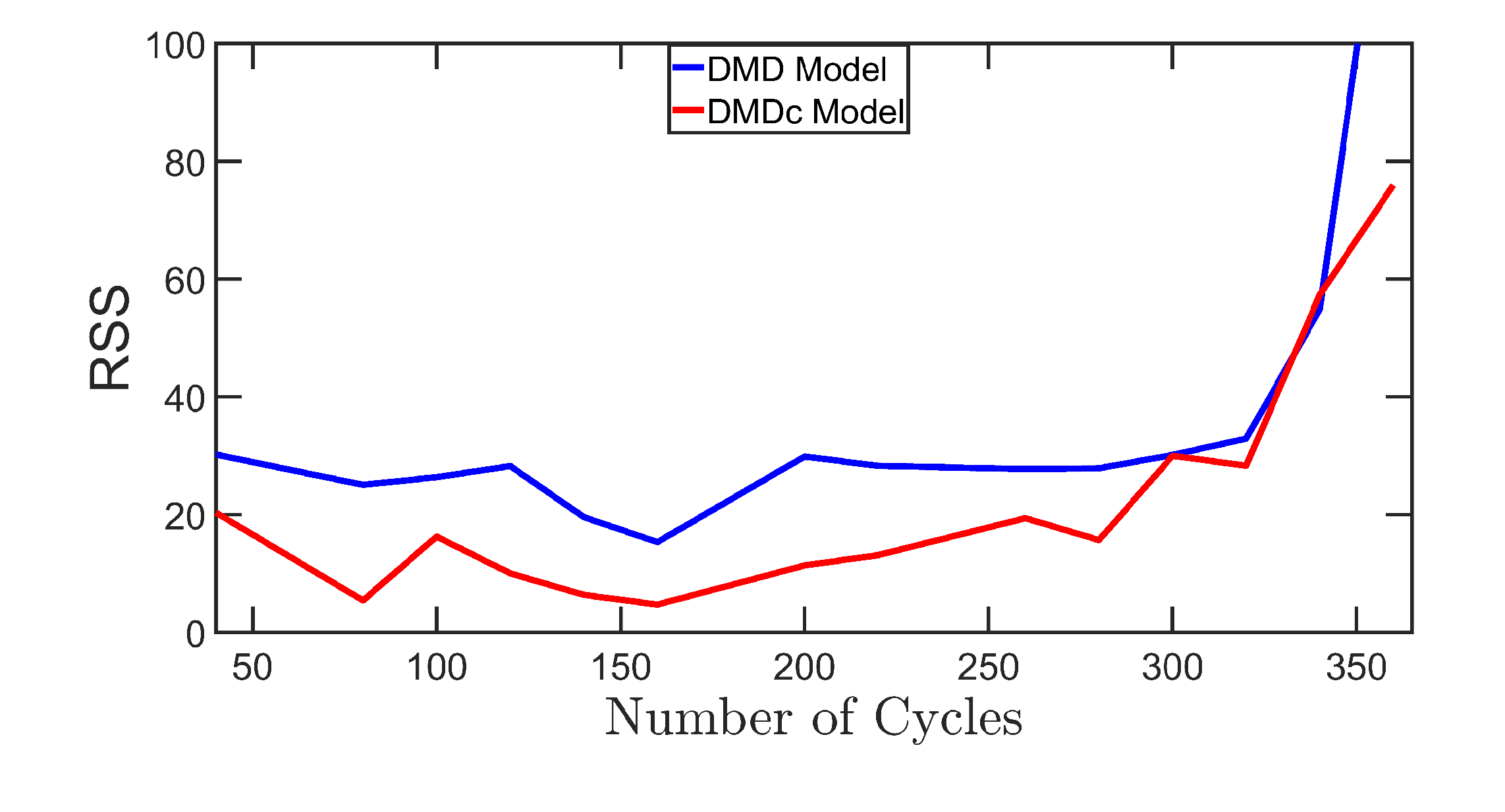}
        
    \end{subfigure}
    
    \caption{Residual sum of squares (RSS) values at different degradation levels of the battery obtained from the identified DMD and DMDc model. }
    \label{fig:RSS}
\end{figure}

We swept the embedding dimension \(m\in[1000,1900]\) and computed the residual sum of squares (RSS) for each case. The lowest residual (31\%) for plain DMD (no inputs) occurred at \(m=1810\) (\cref{fig:RSS1}a). Incorporating the known current as an exogenous input via DMD with control (DMDc), i.e., \(\mathbf{X}' \approx \mathbf{A}\mathbf{X}+\mathbf{B}\mathbf{U}\), substantially improved accuracy, reducing the residual to 3.85. To further enhance performance, we also delay-embedded the input to form \(\mathbf{U}=\mathcal{H}_{\ell}(u)\) and tested \(\ell=1,\dots,12\); a six-delay input embedding yielded the lowest residual of 1.74 (\cref{fig:RSS}b).

 \subsection{Model Identification and Validation}
 

For model identification, the first \(60\%\) of the voltage and current samples from each cycle were used for training, with the remainder reserved for evaluation. The system matrices were estimated in least squares as described earlier: the DMD model yields \(\mathbf{A}\) from \(\mathbf{X}'\approx \mathbf{A}\mathbf{X}\), and the DMDc model yields \([\mathbf{A}\;\mathbf{B}]\) from \(\mathbf{X}'\approx \mathbf{A}\mathbf{X}+\mathbf{B}\mathbf{U}\).
After estimating \(\mathbf{A}\) (and \(\mathbf{B}\) for DMDc) and specifying the initial condition, we rolled the model forward in time via
\(\mathbf{x}_{k+1}=\mathbf{A}\mathbf{x}_k\) (DMD) or \(\mathbf{x}_{k+1}=\mathbf{A}\mathbf{x}_k+\mathbf{B}\mathbf{u}_k\) (DMDc),
treating \(\mathbf{x}_k\) as the delay-embedded (Hankel) state. This produced multi-step, open-loop forecasts that were compared against experimental measurements.

\begin{figure}[t!]
    \centering
    \begin{picture}(200,290)
    \put(-20,150){ \includegraphics[width=0.9\columnwidth]{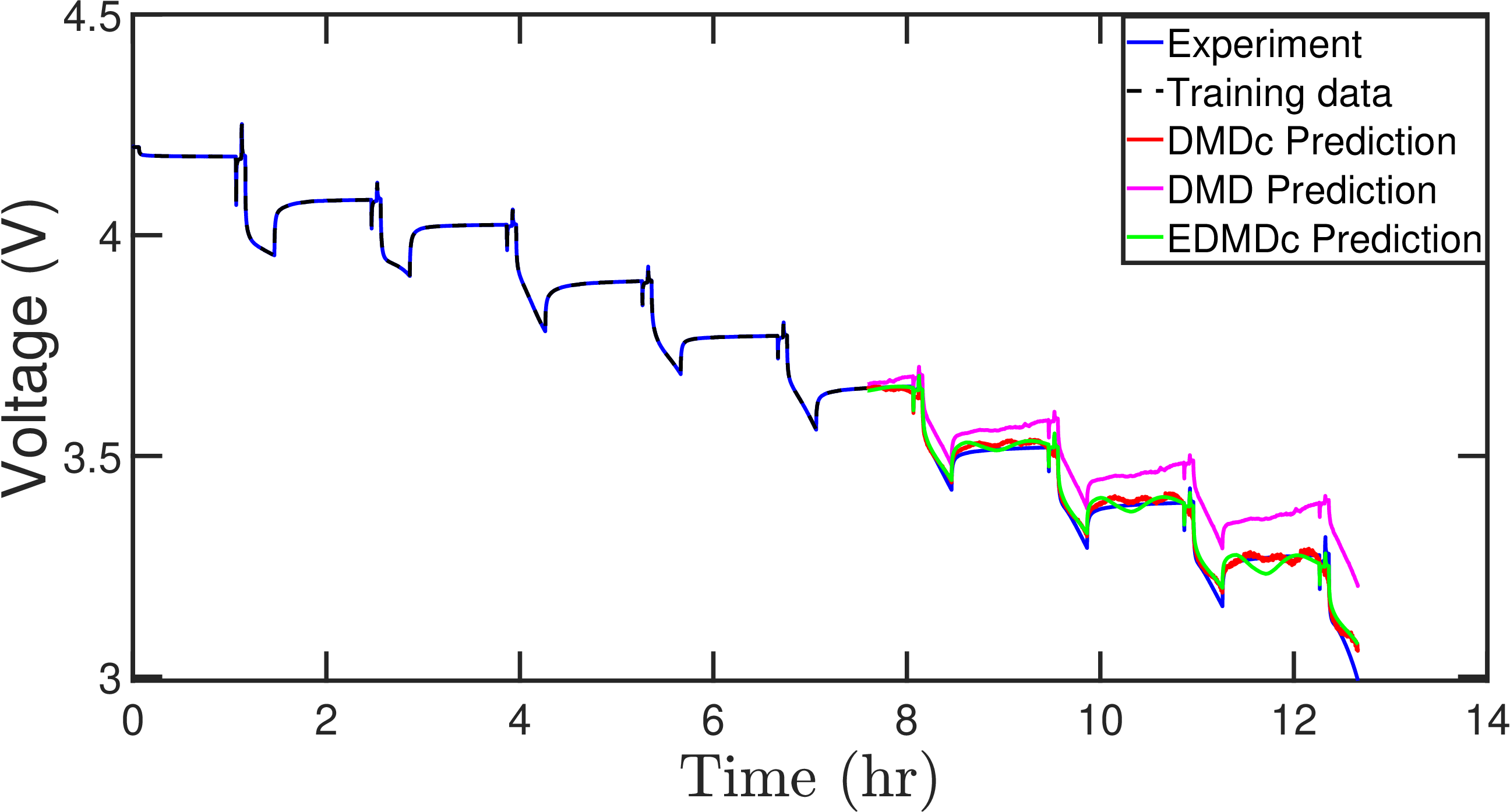}}
    \put(-20,15){\includegraphics[width=0.9\columnwidth]{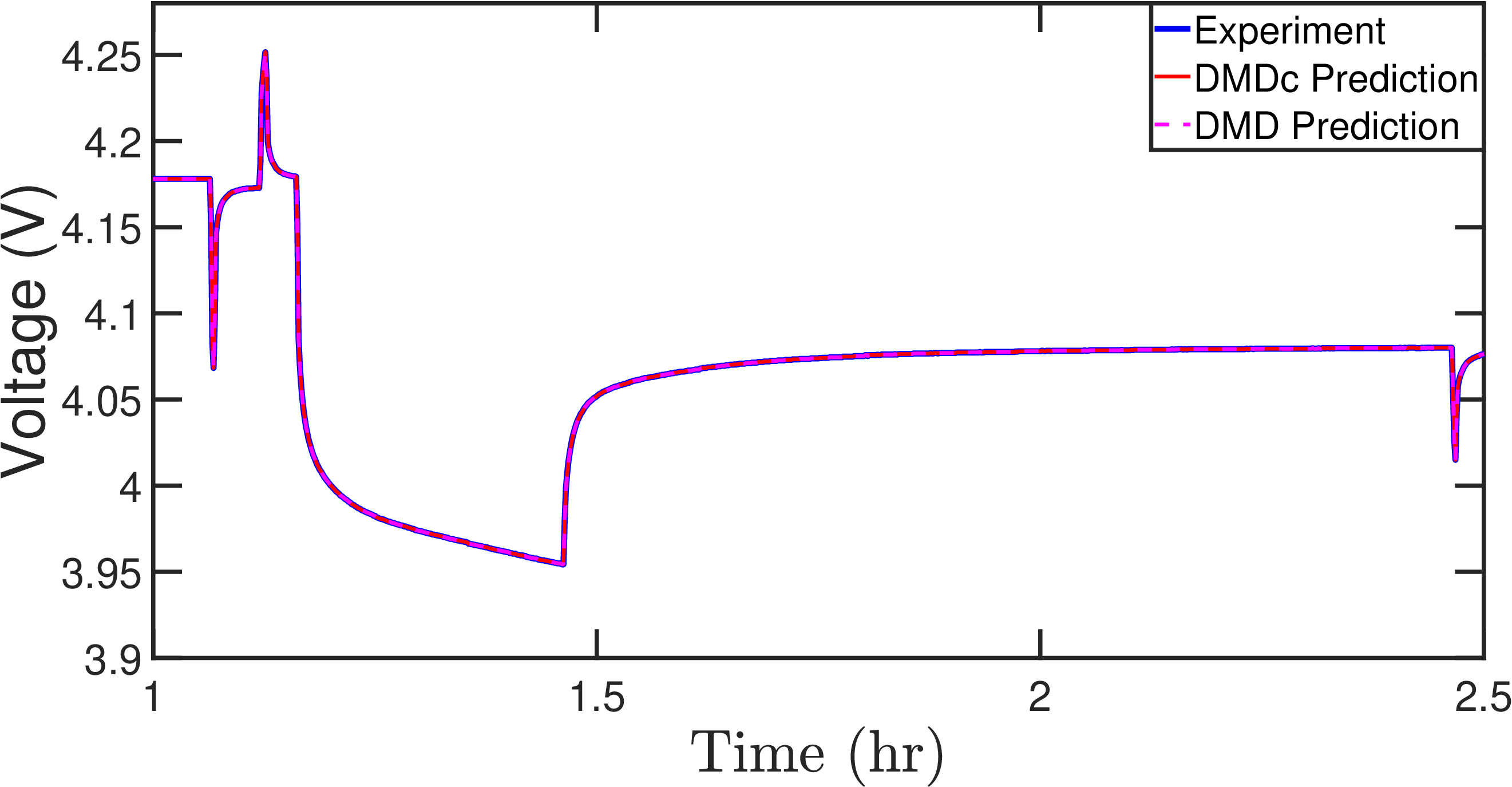}}
    \put(-20,-120){\includegraphics[width=0.9\columnwidth]{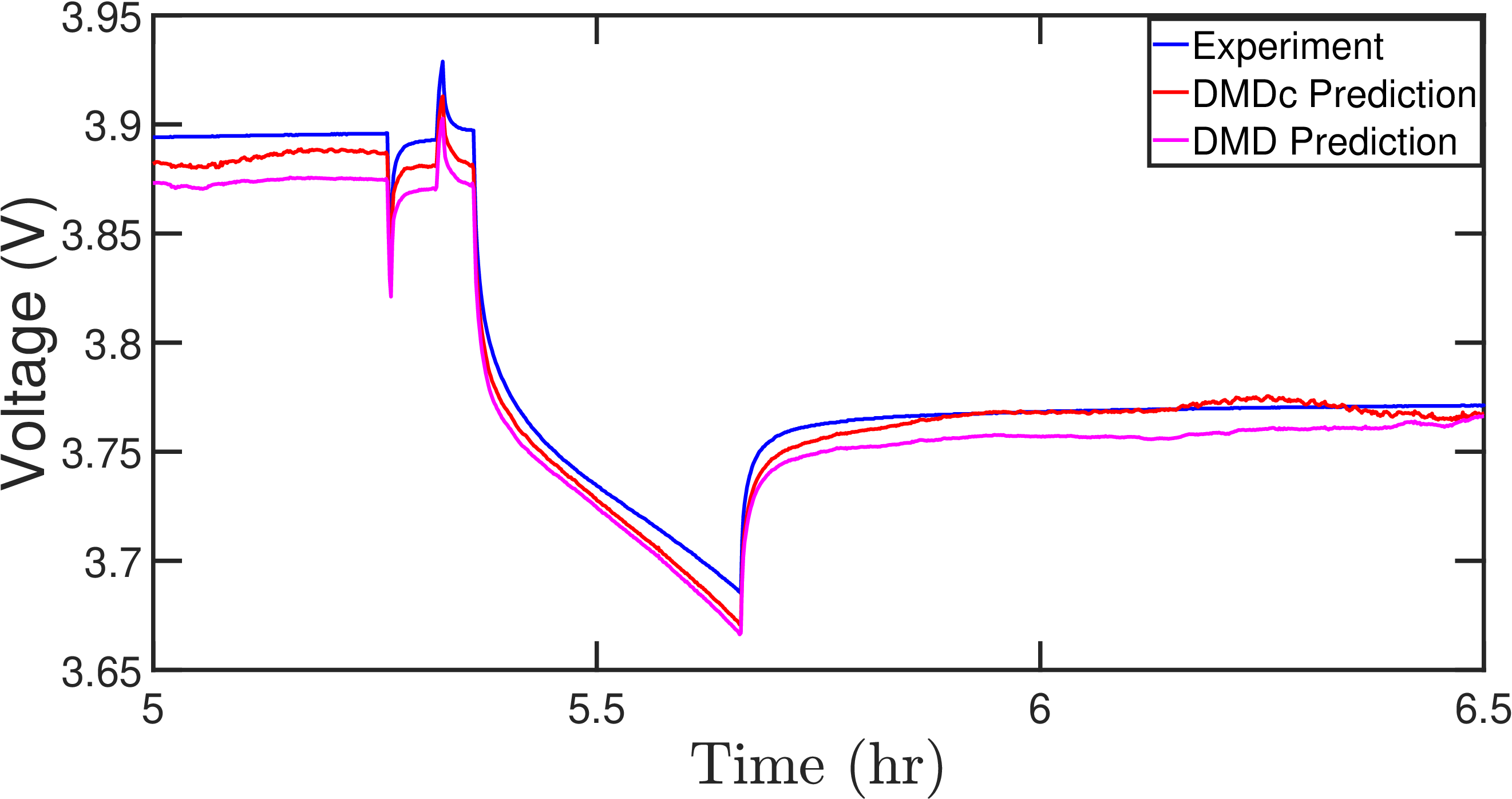}}
    \put(-20,-255){\includegraphics[width=0.9\columnwidth]{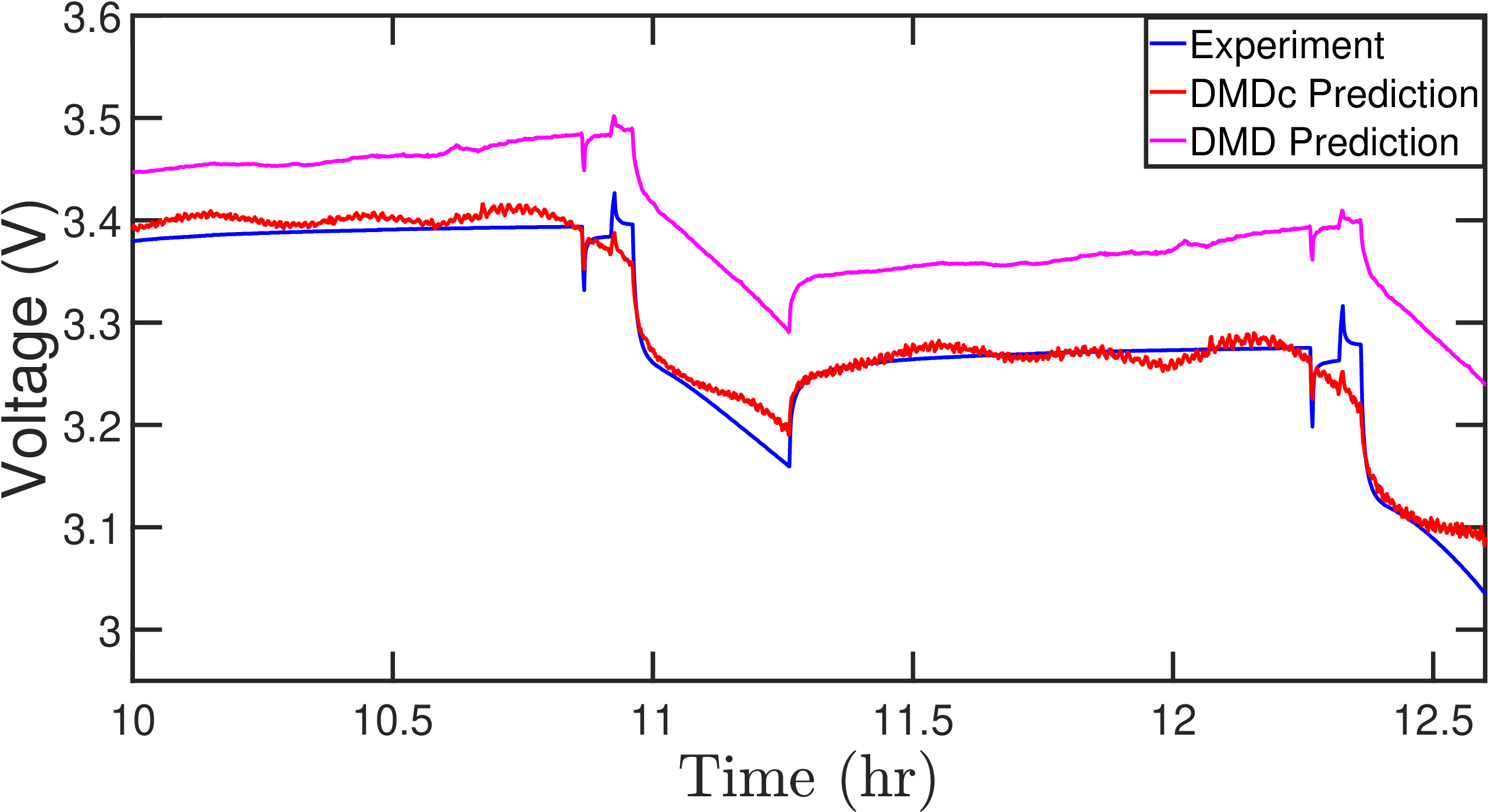}}
    
    \put(10,255){\large \textbf{(a)}}
    \put(15,105){ \large \textbf{(b)}}
    \put(10,-20){\large \textbf{(c)}}  
    \put(10,-150){\large \textbf{(d)}}  
    \end{picture}
    \vspace{9.5cm}
    
    \caption{Comparison of discharge profiles with actual measurements, DMD and DMDc models: (a) shows the complete voltage response curve with training and forecast part, (b) prediction from 1 hour to 2.5 hours; (c) prediction from 5 hours to 6.5 hours; (d) prediction from 10 hours to 12.5 hours.} 
\label{fig:Experiment to model} 
\end{figure}
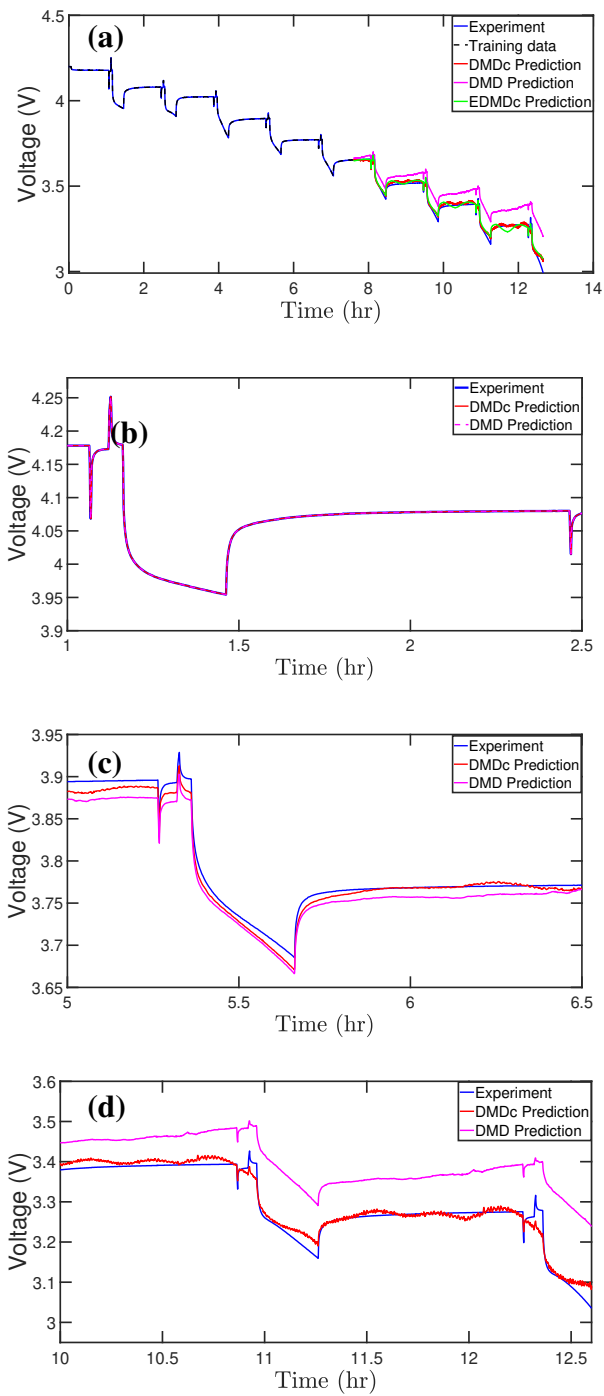

Both models reproduced the pulse-response characteristics of the charge-discharge profiles (Figure \ref{fig:Experiment to model}(a)), with DMDc consistently attaining higher accuracy. In the initial stage (up to \(\sim 2.5\,\mathrm{h}\)), predictions from both models closely matched the measurements (Figure \ref{fig:Experiment to model}(b)). As the forecast horizon increased, discrepancies grew more visible: the standard DMD model incurred larger errors beyond \(\sim 6.5\,\mathrm{h}\), whereas the DMDc model maintained accurate pulse-response predictions with negligible deviation (\ref{fig:Experiment to model}(c)). In the late stage (after \(\sim 10\,\mathrm{h}\)), the DMD model showed substantially larger deviations, although it still captured the intrinsic discharge trend (\ref{fig:Experiment to model}(d)). The DMDc model demonstrated strong long-horizon performance, achieving RSS of 1.74 even for far-future predictions.

\subsection{Adaptability of Models at Degraded States }

\begin{figure}[t!]
       \centering
    \begin{picture}(200,290)
    \put(-20,150){ \includegraphics[width=\columnwidth]{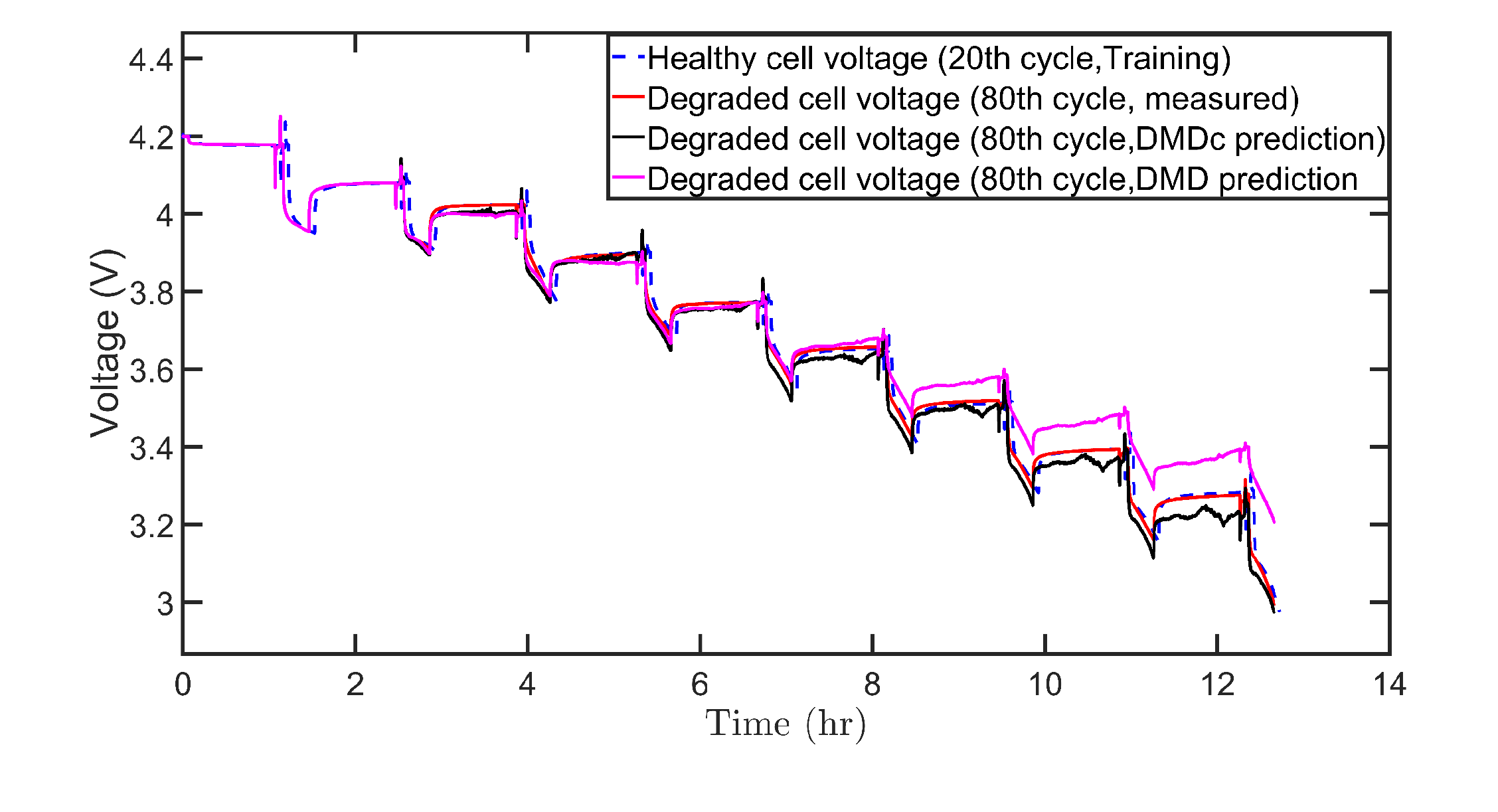}}
    \put(-10,5){\includegraphics[width=\columnwidth]{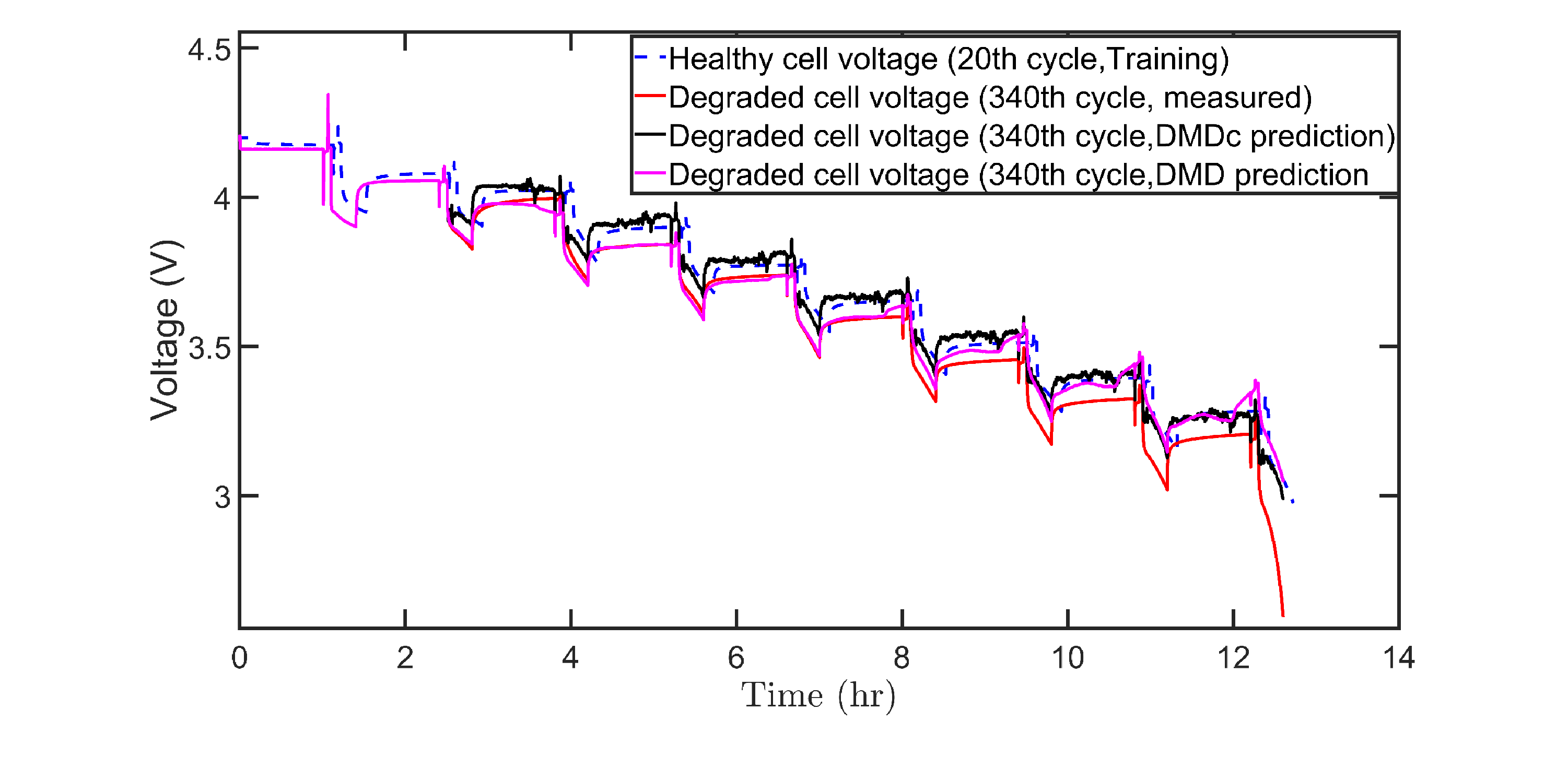}}

    \put(30,250){\large \textbf{(a)}}
    \put(30,105){ \large \textbf{(b)}}
    \end{picture}
    \vskip\baselineskip

    \caption{Comparison of experimental measurements with DMDc model predictions, where the model was trained using data from the 20th cycle: (a) prediction for the 80th cycle, and (b) prediction for the 340th cycle.}
    \label{fig:degraded_model}
\end{figure}

The identified model was applied to HPPC datasets collected at different degradation stages of the battery, as determined by the number of charge-discharge cycles. Consistent delay-embedding parameters were used across all cases to ensure comparability. Both the dynamic mode decomposition (DMD) and its control-augmented counterpart (DMDc) successfully reproduced the pulse-response behavior of the cells. While the DMD model produced residual sum of squares (RSS) values ranging from 30 to 143 across various state of charge (SoC) and degradation levels (\cref{fig:degraded_model}), the DMDc model achieved markedly superior predictive performance, with RSS values between 20 and 80. At cycle 360, which had a higher level of degradation, the RSS rose to 143 for DMD and 76 for DMDc (\cref{fig:degraded_model}). Overall, these results demonstrate the robustness of the DMDc framework over the standard DMD approach in forecasting and simulating battery charge-discharge dynamics across different stages of degradation and SoC.


\section{Conclusions} 




In this work, dynamic mode decomposition (DMD) and its controlled variant (DMDc) were applied to model the dynamical behavior of lithium-ion batteries. Unlike conventional model-based approaches that rely on Equivalent Circuit Models (ECMs) or electrochemical partial differential equations, and unlike data-driven approaches that require derived features such as state of charge (SoC) or capacity, our framework directly utilizes raw time-series voltage responses. A Hankel matrix embedding was constructed from sequential voltage data, which served as the state matrix for DMD and DMDc. This embedding lifted the original signal into a higher-dimensional space, where latent dynamical patterns associated with nonlinear charge–discharge processes were represented through an approximate linear evolution operator. This provides a systematic way of capturing nonlinear dynamics within an interpretable, linearized state-space framework.

The models were evaluated under different state of charge and health conditions of the battery using hybrid pulse power characterization (HPPC) test data. Initially, the $A$ (and $B$) matrix was identified with the 60\% of the available HPPC test data collected at the battery's healthy state, and the simulation was performed for the entire data set using both DMD and DMDc model. The simulation successfully reproduced the full charge–discharge voltage profile with minimal error. The DMDc model consistently achieved higher predictive accuracy, with a residual sum of squares (RSS) as low as 1.29, highlighting the benefit of explicitly accounting for control input (current) in the decomposition. More importantly, the adaptability of the DMD and DMDc framework was demonstrated by identifying the $A$ (and $B$) matrix solely from the healthy-state HPPC data and then applying the learned models to simulate degraded-cell responses. Both DMD and DMDc reproduced the voltage response of the degraded batteries, with DMDc showing greater robustness in capturing the altered dynamics induced by degradation.

The novelty of this work lies in modeling the non-linear dynamics of a battery with linear and interpretable models. By employing Hankel embeddings of raw voltage signals, the method circumvents the need for difficult-to-measure internal states or handcrafted degradation indicators, while still yielding interpretable linear state-space models. This positions DMD and DMDc as powerful alternatives to black-box machine learning models, which often lack transparency, and to ECMs, which may oversimplify nonlinearities. The results highlight the potential of DMD-based methods to form the foundation of adaptive battery diagnostics and prognostics frameworks, capable of tracking evolving nonlinear behavior in real time.

Overall, the study demonstrates that DMDc, in particular, offers superior predictive performance and robustness, while both DMD and DMDc provide interpretable linear models for complex nonlinear dynamics. This approach opens new directions for battery state estimation and degradation modeling, with the potential to extend beyond lithium-ion systems to other energy storage technologies.

\end{document}